\documentclass[iop,apj]{emulateapj}
\usepackage{graphicx,url,amssymb,amsmath,longtable,rotating,color,units,wasysym,subfigure,epsfig,multirow,epstopdf,enumerate,tabularx}
\usepackage[colorlinks,urlcolor=blue,citecolor=blue,linkcolor=blue]{hyperref}

\newcommand{\heavi}[1]{\mathcal{H}(#1)}
\newcommand{\msol}[1]{\unit{#1}{M_{\odot}}}
\newcommand{\al}{\alpha}
\newcommand{\mx}{m_{\max}}
\newcommand{\mn}{m_{\min}}
\newcommand{\lam}{\lambda}
\newcommand{\mpp}{m_{\text{pp}}}
\newcommand{\spp}{\sigma_{\text{pp}}}
\newcommand{\bt}{\beta}
\newcommand{\dm}{\delta m}
\newcommand{\mzams}{M}

\usepackage{lineno}

\begin{document}

\title{Measuring the binary black hole mass spectrum \\with an astrophysically motivated parameterization}

\author{Colm Talbot}
\email{colm.talbot@monash.edu}
\author{Eric Thrane}
\affiliation{School of Physics and Astronomy, Monash University, Clayton, Victoria 3800, Australia}
\affiliation{OzGrav: The ARC Centre of Excellence for Gravitational-wave Discovery, Clayton, Victoria 3122, Australia}


\begin{abstract}

Gravitational-wave detections have revealed a previously unknown population of stellar mass black holes with masses above $\msol{20}$.
These observations provide a new way to test models of stellar evolution for massive stars.
By considering the astrophysical processes likely to determine the shape of the binary black hole mass spectrum, we construct a parameterized model to capture key spectral features that relate gravitational-wave data to theoretical stellar astrophysics.
In particular, we model the signature of pulsational pair-instability supernovae, which are expected to cause all stars with initial mass $\msol{100}\lesssim M \lesssim \msol{150}$ to form $\sim \msol{40}$ black holes.
This would cause a cut-off in the black hole mass spectrum along with an excess of black holes near $\msol{40}$.
We carry out a simulated data study to illustrate some of the stellar physics that can be inferred using gravitational-wave measurements of binary black holes and demonstrate several such inferences that might be made in the near future.
First, we measure the minimum and maximum stellar black hole mass.
Second, we infer the presence of a peak due to pair-instability supernovae.
Third, we measure the black hole mass ratio distribution.
Finally, we show how inadequate models of the black hole mass spectrum lead to biased estimates of the merger rate and the amplitude of the stochastic gravitational-wave background.
\end{abstract}

\maketitle

\section{Introduction} \label{sec:intro}
The black holes observed by advanced gravitational-wave detectors such as the Laser Interferometer Gravitational-wave Observatory (LIGO) (\cite{aLIGO}) and Virgo (\cite{AdVirgo}) are widely believed to be formed from massive stars with initial mass, $\mzams\gtrsim\msol{20}$ (\cite{Heger2003}).
Gravitational-wave measurements constrain the mass and spin of merging binaries.
These measurements, in turn, can be used to better understand the evolution of massive stars.
In this paper, we study how gravitational-wave measurements of the black hole mass spectrum can be used to inform our understanding of stellar evolution.

Simulating the final stages of stellar binary evolution is computationally expensive.
Additionally, there are significant theoretical uncertainties in key aspects of binary evolution, especially the common envelope phase (\cite{Ivanova2013}) and supernova mechanism.
For these reasons, populations of compact objects are simulated using population synthesis models (e.g.,~\cite{Dominik2015,Belczynski2017,Stevenson2017a}).
These are phenomenological models calibrated against a small number of more detailed stellar simulations.

At the time of writing, there have been five confirmed detections of binary black hole mergers and one unconfirmed candidate event (\cite{GW150914,PhysRevLett.116.241103,PhysRevX.6.041015,GW170104,GW170814,GW170608}).
The 90\% credible regions for the source masses of the black holes range from $\sim \msol{5}$ to $\sim \msol{40}$.
Of the observed events, only one (GW151226) provides unambiguous evidence of black hole spin (\cite{PhysRevLett.116.241103}), although two events (GW150914 and GW170104) show a weak preference for spins anti-aligned with respect to the orbital angular momentum vector (\cite{PhysRevD.93.122003,GW170104}).
The implications of these measurements are currently unclear.
Current theories include: most binary black holes are formed dynamically (\cite{Rodriguez2016b,Rodriguez2017}), black holes are subject to large black hole natal kicks (\cite{Rodriguez2016a,OShaughnessy2017}), and/or that large black holes do not form with significant dimensionless spins (\cite{Belczynski2017,Wysocki17}).

There has been significant work using gravitational-wave data to infer the properties of black hole formation with ensembles of detections.
These works range from comparing gravitational-wave data to specific, non-parameterized models (\cite{Mandel2010,Stevenson2015,Dominik2015,Belczynski2016b,Stevenson2017b,Zevin2017,Belczynski2017,Miyamoto17,FarrW17,Wysocki17,Barrett2017}), to attempts to group the data by binning, clustering or Gaussian mixture modeling (\cite{Mandel2017,FarrB17,WysockiThesis}), to fitting physically motivated phenomenological population (hyper)parameters (\cite{Kovetz2017a,Talbot17,Fishbach17b}).
In this work, we take the last approach and demonstrate that it is possible to identify physical features in the black hole mass spectrum with an ensemble of detections using phenomenological models, building on work in \cite{Kovetz2017a} and \cite{Fishbach17b}.

Previous attempts to determine the binary black hole mass spectrum have employed one or more of these three approaches.
Clustering is applied to a binned mass distribution in \cite{Mandel2017} to demonstrate that a mass gap between neutron star and black hole masses can be identified after $O(100)$ observations.
In (\cite{Zevin2017,Stevenson2015,Barrett2017}), the authors compare population synthesis models with different physical assumptions and show that predicted mass distribution can be distinguished using $O(10)$ of observations.

Previous analyses by the LIGO and Virgo scientific collaborations fit a power-law model with variable spectral index, $\alpha$.
\citeauthor{Fishbach17b} point out that, given LIGO/Virgo's additional sensitivity to heavier binary systems, there is a possible dearth of black holes larger than $\sim\msol{40}$.
They suggest that this is due to the occurrence of pulsational pair-instability supernovae (\cite{Heger2002}) and propose an extension of the current LIGO analysis where the maximum mass is a free parameter.
Pulsational pair-instability supernovae occur in stars with initial masses $\msol{100} \lesssim M \lesssim \msol{150}$, causing all stars in that mass range to form black holes with mass $\sim \msol{40}$.
In addition to a cut-off in the black hole mass spectrum, we expect that there will be an excess of black holes around the cut-off mass.

\citeauthor{Kovetz2017a} model the lower mass limit of black holes and use a Fisher analysis to demonstrate that it should be possible to identify the presence of the neutron-star black hole mass gap.
They also model the distribution of mass ratios and propose a test for detecting primordial black holes.
A method to simultaneously estimate the binary black hole mass spectrum and the merger rate is presented in~\cite{WysockiThesis}.
\citeauthor{WysockiThesis} also considers a Gaussian mixture model for fitting the distribution of compact binary parameters.

The rest of the paper is structured as follows.
In section~\ref{sec:inference}, we introduce the statistical tools necessary to make statements about the black hole population.
We then develop our model in section~\ref{sec:phenom} in terms of population (hyper)parameters by considering current observational constraints and predictions from theoretical astrophysics and population synthesis.
In section~\ref{sec:mc}, we perform a Monte Carlo injection study.
We consider how many detections will be necessary to identify different features using Bayesian parameter estimation and model selection.
We show how the predicted mass distributions differ when using different (hyper)parameterizations.
We also explore some of the consequences of using inadequate (hyper)parameterizations.
In particular, we show that inadequate (hyper)parameterization can lead to significant bias in the estimate of the merger rate and the predicted amplitude of the stochastic gravitational-wave background (SGWB).
Some closing thoughts are provided in section~\ref{sec:discussion}.

\section{Bayesian Inference} \label{sec:inference}

\subsection{Gravitational-wave Detection}

A binary black hole system is completely described by  15 parameters, $\Theta$.
Recovering these parameters from the observed strain data requires the use of specialized Bayesian parameter inference software, e.g., \texttt{LALInference} (\cite{PhysRevD.91.042003}).
The likelihood of a given set of binary parameters is computed by comparing the strain data to the signal predicted by general relativity.
For the analysis presented here, the expected signal is calculated using phenomenological approximations to numerical relativity waveforms (\cite{Hannam2014,Schmidt2015,Smith2016}).
For a given set of strain data, $h_i$, \texttt{LALInference} returns a set of $n_i$ samples, $\{\Theta\}$, which are sampled from the posterior distribution, $p\left(\Theta\middle|h_i, H\right)$, of the binary parameters, along with the Bayesian evidence, $\mathcal{Z}(h_i|H)$, where $H$ is the model being tested.

The distribution of binary black hole systems observed by current detectors is not representative of the astrophysical distribution of binary black holes.
The observing volume of current gravitational-wave detectors is limited by the instruments' sensitivity.
The sensitive volume for a detector to a given binary is primarily determined by the masses of the black holes with spin entering as a higher order effect.
More massive systems produce gravitational waves of greater amplitude.
However, these more massive systems merge at a lower frequency and, hence, spend less time in the observing band of the detector.
Additionally, distant sources undergo cosmological redshift and appear more massive than they actually are.
Here, we will deal only with the un-redshifted ``source-frame'' masses, not the ``lab-frame'' masses observed by gravitational-wave detectors
We note that the source/lab-frame distinction is about cosmological redshift and is not a statement about detectability and/or selection effects.

Accounting for these factors, we calculate $V_{obs}(\Theta)$, the sensitive volume for a binary with parameters $\Theta$, following~\cite{PhysRevX.6.041015}, using semi-analytic noise models corresponding to different sensitivities (\cite{Abbott2016}).
The noise, and hence sensitivity, in real detectors is time-dependent and so calculating this volume requires averaging over the observing time to obtain a mean sensitive volume $\langle V_{obs}(\Theta) \rangle$ (\cite{PhysRevX.6.041015}).

\subsection{Population Inference}

We are interested in inferring population (hyper)parameters describing the distribution of source-frame black hole masses.
The formalism to do this is briefly described below (see e.g., \cite{gelman2013bayesian}, chapter 29 for a more detailed discussion of hierarchical Bayesian modeling and~\cite{Mandel2014} for a discussion of selection biases).

Hierarchical inference of this kind can be cast as a post hoc method of changing from the prior distribution used in the single event parameter estimation to a new prior, which depends on population (hyper)parameters, $\Lambda$.
We marginalise over all of the binary parameters while reweighting the posterior samples by the ratio between our (hyper)parameterized model and the prior used to generate the posterior distribution.
This marginalisation integral is approximated by summing over the posterior samples for each event.
The $N$ events are then combined by multiplying the new marginalised likelihood for the individual events,
\begin{equation}
    \mathcal{L}\left(\{h\}_{i=1}^{N}\middle|\Lambda, H \right) \propto \prod_{i}^{N} \sum^{n_i}_{j} \frac{\pi(\Theta^i_j|\Lambda, H)}{\pi(\Theta^i_j|\text{LAL})} .
\label{eq:hyperlike}
\end{equation}
Here $\pi(\Theta|\text{LAL})$ is the prior probability distribution used for single event parameter estimation.
The distribution $\pi(\Theta^i_j|\Lambda, H)$ is the probability of a binary having parameters $\Theta$ in our model; see Sec.~\ref{sec:phenom}.
We do not model any black hole parameters other than source-frame mass and so $\Theta$ can be replaced by $m=(m_1, m_2)$, in Eq.~\ref{eq:hyperlike} and all following equations.
The prior distribution of source-frame masses in \texttt{LALInference} is a convolution of a uniform in component mass prior between limits determined by the reduced order model~\cite{Smith2016} and the redshift distribution corresponding to uniform in luminosity distance extending out to $\unit{4}{Gpc}$.
This distance distribution is converted to redshift assuming $\Lambda$CDM cosmology using the results from the \textit{Planck} 2015 data release (\cite{Planck15}).

We combine this likelihood with $\pi(\Lambda| H)$, the prior for the (hyper)parameters assuming a model $H$, and the Bayesian evidence for the data given $H$ to obtain the posterior distribution for our (hyper)parameters,
\begin{equation}
p\left(\Lambda\middle|\{h\}_{i=1}^{N}, H \right) =
\frac{\mathcal{L}\left(\{h\}_{i=1}^{N}\middle|\Lambda, H \right)\pi(\Lambda| H)}{\mathcal{Z}\left(\{h\}_{i=1}^{N} \middle|H \right)},
\end{equation}
\begin{equation}
\mathcal{Z}\left(\{h\}_{i=1}^{N} \middle|H \right) = \int d\Lambda \, \mathcal{L}\left(\{h\}_{i=1}^{N}\middle|\Lambda, H \right)\pi(\Lambda| H).
\label{eq:ev}
\end{equation}

To perform the (hyper)parameter estimation we use the python implementation of \texttt{MultiNest} (\cite{Feroz09,Buchner14}).
Additionally, we calculate the posterior predictive distribution (PPD) of the binary parameters,
\begin{align}
p(m|\{h\}_{i=1}^{N}, H)
&= \int \, d\Lambda \pi(m|\Lambda, H) p(\Lambda|\{h\}_{i=1}^{N}, H) \nonumber \\
&\approx \frac{1}{n_{k}} \sum_{k}^{n_{k}} \pi(m|\Lambda_{k}, H),
\label{eq:ppd}
\end{align}
where $\Lambda_{k}$ are the $n_{k}$ (hyper)posterior samples.
The PPD shows the probability that a subsequent detection will have parameters $\Theta$ given the previous data, $\{h\}$.

\subsection{Model Selection}

Model selection is performed in our Bayesian framework by considering Bayes factors,
\begin{equation}
BF^{\al}_{\bt} = \frac{\mathcal{Z}\left(\{h\}_{i=1}^{N} \middle|H_{\al} \right)}{\mathcal{Z}\left(\{h\}_{i=1}^{N} \middle|H_{\bt} \right)}.
\end{equation}
A large Bayes factor, $BF^{\al}_{\bt} \gg 1$, indicates that $H_{\al}$ is strongly favored over $H_{\bt}$.
We adopt a conventional threshold of $\ln BF=8$ to distinguish between two models.

\section{Phenomenology} \label{sec:phenom}

In this section, we develop a parameterization of the black hole mass spectrum using predictions from astrophysics theory, population synthesis models, and electromagnetic observations.
In this way, we can relate gravitational-wave measurements to stellar astrophysics.
For low-mass systems, the parameter that most strongly affects the observable gravitational waveform is a combination of the component masses known as the chirp mass, $\mathcal{M}=(m_1m_2)^{3/5}/(m_1+m_2)^{1/5}$.
For high-mass systems, the waveform is primarily determined by the total mass of the system.
The mass ratio is more difficult to determine due to covariances between the mass ratio and the spin of the black holes.

The canonical assumed distribution of black hole masses is a power law distribution in the primary mass between some maximum and minimum masses.
This power law distribution has three typical parameters: the spectral index $\al$, the minimum mass $\mn$, and the maximum mass $\mx$.
The distribution of secondary mass is typically taken to be flat between $\mn$ and $m_{1}$.
We take this as the starting point for our parameterization.

\subsection{High-Mass Binaries}

The observation of binary black hole mergers through the detection of gravitational waves revealed the presence of a previously unobserved population of black holes with mass $\sim \msol{30}$ (\cite{GW150914, GW170104, GW170814}).
Since gravitational-wave detectors can observe more massive binaries at greater distances, binaries containing larger black holes are preferentially detected over less massive systems.
\citeauthor{Fishbach17b} note that, given the observation rate of binaries with mass $\sim \msol{30}$, it is somewhat surprising that we have not seen more massive black holes.
They propose that this is due to a cut-off in the black hole mass spectrum around this mass.
By comparing the Bayesian evidence for $\mx=\msol{41}$ and $\mx=\msol{100}$ using the first four events, they find tentative support for a cut-off.
They further show that it will be possible to identify the presence of an ``upper mass gap'' with a Bayes Factor of $\gtrsim 150$ ($\ln BF \approx 5$)  using 10 detections and the cut-off mass can be measured with $\sim 40$ detections.

The theoretical motivation for such a cut-off is pulsational pair-instability supernovae (PPSN) (\cite{Heger2002, Woosley2015}), whereby large amounts of matter are ejected prior to collapse to form a black hole.
The expected result of this process is that all stars with initial mass $\msol{100}\lesssim \mzams\lesssim\msol{150}$ form black holes with masses $\sim\msol{40}$.
Stars with $\msol{150}\lesssim \mzams\lesssim\msol{250}$ are expected to undergo pair-instability supernovae (PISN) and leave no remnant.
Hence, we expect a gap in the black hole mass spectrum between $\sim\msol{40}$ and $\sim\msol{250}$ along with an excess of black holes at some mass $\mpp \sim \msol{40}$.
The excess is due to the $\msol{100}\lesssim \mzams\lesssim\msol{150}$ stars which undergo PPSN.
The size, position and shape are determined by the unknown details of PPSN.
While the observation of cut-off near $\msol{40}$ could be interpreted as evidence for PPSN, the additional observation of a peak would provide a smoking-gun signature that the highest mass stellar binaries are reduced in mass via PPSN.

Thus, we extend our description of the upper end of the black hole mass spectrum to allow for the possibility of an excess due to PPSN.
We model this as a normal distribution with unknown mean $\mpp\in [25, 100] \msol{}$ and variance $\spp < \msol{5}$.
It is also necessary to introduce a mixing fraction parameter, $\lam$, which describes the proportion of binaries which are drawn from the normal distribution.

We expect that any PPSN mass peak should be near the high-mass cut-off, this corresponds to $m_{\max}\approx\mpp$.
Also, assuming that the power law is otherwise a good description of the black hole mass spectrum, we expect that the number of black holes in the PPSN peak should be no more than the number of black holes that would have formed had the power law distribution continued to the upper limit of the upper mass gap, determined by the onset of pair-instability supernovae, $m_{\text{PI}} \sim \msol{150}$.
We impose this condition by requiring that the extrapolated area which would be under the power-law curve is less than the area contained within the Gaussian.
This amounts to a restriction on the allowed values of $\lambda$,
\begin{equation}
\lam \leq \frac{\int_{m_{\max}}^{m_{\text{PI}}}m^{-\alpha}}{\int_{m_{\min}}^{m_{\text{PI}}}m^{-\alpha}} =
 \frac{m_{\text{PI}}^{1-\al}-\mx^{1-\al}}{m_{\text{PI}}^{1-\al}-\mn^{1-\al}} \approx \left(\frac{\mn}{\mx}\right)^{\al-1} .
\end{equation}
Here, $\mn$ is the upper limit of the NS-BH mass gap and $\mx$ is the lower limit of the upper mass gap.
The variable $m_{\text{PI}}$ is the mass above which stars undergo PISN leaving no remnant.
Here, we assume $\al>1$ and $m_{\text{PI}}\gg\mx$.

To hone our intuition, we can plug in plausible values of $\al$, $\mn$, $\mx$ and $m_{\text{PI}}$ to determine a typical value of $\lam$.
For example, if we set $\mn=\msol{5}, \mx=\msol{50}$ and $\al=2$, we find $\lam \sim 0.1$.
If we measure a peak consistent with these values, we anticipate that the position and width of the peak can inform our physical understanding of this mechanism.
If $\lam$ is measured to be inconsistent with this constraint it could indicate that either the extrapolation of the power law is not a valid assumption, or that the peak is not entirely due to PPSN.
We note that the fraction of \textit{observed} black holes which formed through PPSN will be larger than $\lam$ since the more massive black holes are observable out to a greater distance.

\subsection{Low-Mass Binaries}

The smaller sensitive volume for lower-mass binaries means that it is more difficult to probe the low-mass end of the black hole mass spectrum with gravitational-wave detections.
Previous analyses of the black hole mass spectrum from gravitational-wave detections have assumed that the black hole mass spectrum has a sharp cut-off at some minimum mass $\mn$.
However, this overestimates the number of low-mass black holes if the distribution of low-mass black holes in merging binaries is the same as that in low-mass X-ray binaries (\cite{Ozel2010}).
Population synthesis models also generically predict that the primary mass distribution peaks above the minimum mass.

We replace the step function at the low-mass end of the black hole mass spectrum with a smoothing function, $S(m, \mn, \dm)$, which rises from zero at $\mn$ to one at $\mn+\dm$,
\begin{equation}
        S(m, \mn, \dm) = \left( \exp f(m-\mn, \dm) + 1 \right)^{-1}
\end{equation}
\begin{equation}
        f(m, \delta m) = \frac{\delta m}{m} - \frac{\delta m}{m - \delta m}.
\end{equation}
We note that $\dm=0$ recovers the step function used in previous analyses.
Since the mass distribution is expected to be an increasing function, the peak of $p(m_1)$ occurs below $\mn+\dm$.
We expect $\mn\sim\msol{5}$ and $\dm \gtrsim 3$ given that the black hole mass spectrum inferred from electromagnetic observations peaks at $\msol{8}$, with no black holes less massive than $\msol{5}$.

Our model of the distribution of the primary mass can be summarized as
\begin{equation}
        p(m_{1}|\Lambda) = (1-\lambda)p_{pow}(m_1|\Lambda) + \lambda \, p_{pp}(m_1|\Lambda)
\label{eq:pm1}
\end{equation}
where
\[
        p_{pow}(m_1|\Lambda) \propto m_{1}^{-\al} S(m_1, \mn, \dm) \heavi{\mx-m_1}
\]
encodes the power-law distribution with a smooth turn on at low mass and
\[
p_{pp}(m_1|\Lambda) \propto \exp\left(-\frac{(m_1-\mpp)^2}{2\spp^2}\right)S(m_1, \mn, \dm).
\]
encodes the peak from PPSN.

\subsection{Mass Ratio}

Previous analyses by the LIGO/Virgo scientific collaborations have assumed that the secondary mass is distributed uniformly between a lower limit set by $\mn$ and an upper limit of $m_{1}$.
This is motivated by observations of the stellar initial binary population, (e.g.,~\cite{Kroupa2013} and~\cite{Belloni18}).
In contrast to this, population synthesis models typically predict that the distribution of mass ratios should be biased towards equal mass binaries, (e.g., \cite{Belczynski2017}).
We model the distribution of the mass ratio as a power-law with spectral index $\beta$ as in~\cite{Kovetz2017a,Fishbach17b}.
For a mass ratio distribution peaked at equal masses, $\bt>0$.
We also impose the same smoothing at the lower limit as we apply to the primary mass.
This allows us to write down the conditional probability distribution for secondary masses given a primary mass,
\begin{equation}
p(m_2|m_1, \Lambda) = \left(\frac{m_{2}}{m_{1}}\right)^{\bt} S(m_2,\mn,\dm) \heavi{m_{1}-m_{2}}.
\label{eq:pq}
\end{equation}

\subsection{Summary}

A table listing the (hyper)parameters and their physical meaning is provided in Tab.~\ref{table:params}.
Including a factor of $V_{obs}$ to account for selection biases, the probability of detecting a mass pair given our (hyper)parameters, $\Lambda=\{\alpha, m_{\min}, m_{\max}, \delta m, \lambda, \mpp, \spp, \beta\}$ and under model $H$, is
\begin{equation}
\pi(m|\Lambda, H) \propto p(m_{1}|\Lambda, H) p(m_2|m_1,\Lambda, H) V_{obs}(m).
\label{eq:model}
\end{equation}

We consider six different models for our (hyper)prior distribution, corresponding to decreasingly stringent physical assumptions as independent hypotheses, these different prior assumptions can be tested with our Bayesian framework using Bayes factors.
In our first model, $H_{0}$, we take the power law distribution with maximum mass, $\mx=\msol{100}$, since the injected data set considered in Sec.~\ref{sec:mc} has a minimum mass of $\msol{3}$, we allow $\mn$ to vary, rather than fixing it at $\mn=\msol{5}$ as in previous analyses.
In $H_{1}$ we introduce a uniform prior on $\mx$.
In order to determine the relative importance of the different features, we switch to models which include all the effects described above except one.
In $H_{2}$, $H_{3}$ and $H_{4}$ we do not include the Gaussian component, mass ratio and low-mass smoothing respectively.
Finally, in $H_{5}$ we include all of these effects.
These prior choices are summarized in Tab.~\ref{table:prior}.

\begin{figure}
\includegraphics[width=\linewidth]{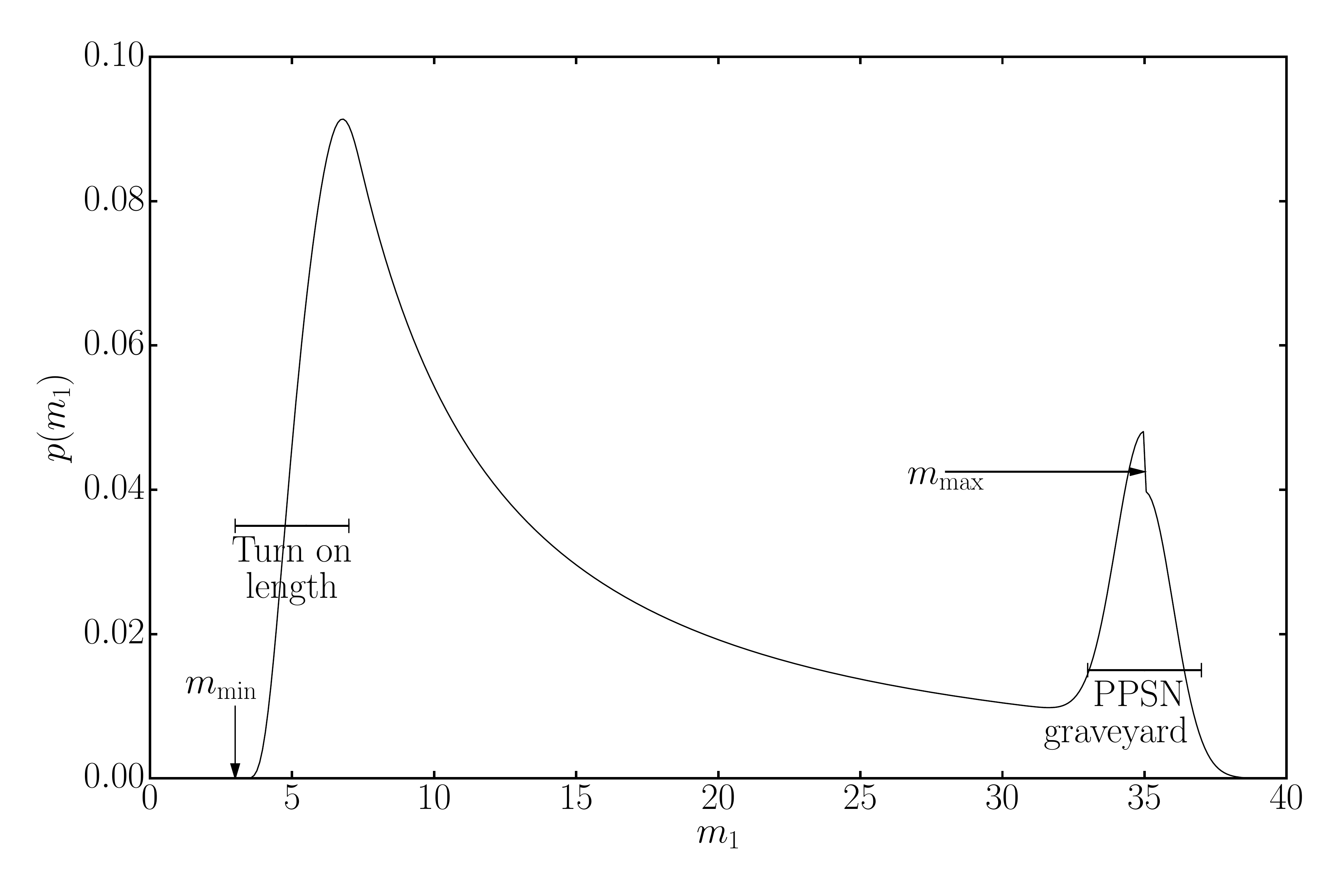}
\caption{The astrophysical distribution of source-frame masses assuming our modeled distribution of $m_{1}$ with (hyper)parameters as specified in Tab.~\ref{table:prior}
We identify the excess of black holes at $\msol{35}$ due to PPSN.
We also see the smooth turn-on at low masses.
}
\label{fig:model}
\end{figure}

\begin{table}[t]
\begin{ruledtabular}
\begin{tabular}{c|l}
$\al$ & Spectral index of $m_1$ for the power-law distributed \\
&  component as the mass spectrum. \\ \hline
$\mx$ & Maximum mass of the power-law distributed \\
&  component as the mass spectrum. \\ \hline
$\lam$ & Proportion of primary black holes formed via PPSN. \\ \hline
$\mpp$ & Mean mass of black holes formed via PPSN. \\ \hline
$\spp$ & Standard deviation of masses of black holes formed \\
& via PPSN. \\ \hline
$\mn$ & Minimum black hole mass. \\ \hline
$\dm$ & Mass range over which black hole mass spectrum \\
& turns on. \\ \hline
$\bt$ & Spectral index of $m_2$. \\
\end{tabular}
\end{ruledtabular}
\caption{(Hyper)parameters describing the black hole mass spectrum.
}
\label{table:params}
\end{table}

\begin{table}[t]
\begin{ruledtabular}
\tabcolsep=0.1cm
\begin{tabular}{c|cc|ccc|c|cc}
 & $\al$ & $\mx$ & $\lam$ & $\mpp$ & $\spp$ & $\bt$ & $\mn$ & $\dm$ \\ \hline
$H_{0}$ & [-3, 7] & \textbf{100} & \textbf{0} & N/A & N/A & \textbf{0} & [2,10] & \textbf{0} \\
$H_{1}$ & [-3, 7] & [10,100] & \textbf{0} & N/A & N/A & \textbf{0} & [2,10] & \textbf{0} \\
$H_{2}$ & [-3, 7] & [10,100] & \textbf{0} & N/A & N/A & [-5,5] & [2,10] & [0,10] \\
$H_{3}$ & [-3, 7] & [10,100] & [0,1] & [25,100] & (0,5] & \textbf{0} & [2,10] & [0,10] \\
$H_{4}$ & [-3, 7] & [10,100] & [0,1] & [25,100] & (0,5] & [-5,5] & [2,10] & \textbf{0} \\
$H_{5}$ & [-3, 7] & [10,100] & [0,1] & [25,100] & (0,5] & [-5,5] & [2,10] & [0,10] \\ \hline
MC & 1.5 & 35 & 0.1 & 35 & 1 & 3 & 5 & 2 \\
\end{tabular}
\end{ruledtabular}
\caption{Summary of example models.
The prior ranges for our (hyper)parameters in each model are indicated.
Each of these distributions is uniform over the stated range.
The fixed parameters are in bold.
``MC'' refers to the values chosen for the simulated universe in Sec.~\ref{sec:mc}.
These values are chosen to be consistent with current observational data and theoretical predictions of pulsational pair-instability supernovae and population synthesis modeling.
}
\label{table:prior}
\end{table}

\subsection{Other Effects}

As with any phenomenological model, our model has limitations.
If the proposed mass gaps exist in the population of black holes formed as the endpoint of stellar evolution there may still be black holes found in these gaps.
The remnant of the binary neutron star merger GW170817 has mass $M_{\text{rem}} \lesssim \msol{2.8}$ (\cite{GW170817}).
It is not clear whether this object is a neutron star or a black hole.
Similarly, the remnant from binary black hole mergers such as GW150914 is more massive than the suggested upper mass limit due to PPSN, $M_{\text{rem}}=\msol{62^{+4}_{-4}}$ (\cite{GW150914PE}).
Both of these objects lie within the proposed mas gaps.
If either of these mergers happened in a dense environment such as a globular cluster, it is possible that such objects could merge with a new companion (\cite{Heggie1975,Rodriguez2017}).
Similarly, primordial black holes (\cite{Hawking71}) are not bound by the limitations of stellar evolution.

We do not expect either of these mechanisms to significantly affect the position and shape of an excess due to PPSN, although they complicate the interpretation of the maximum black hole mass.
It is possible that black holes formed through repeated mergers could be identified on a case by case basis.
For example, black holes formed by a binary black hole merger event are expected to have large dimensionless spins, $a\sim0.7$ for equal mass non-spinning pre-merger black holes (\cite{Scheel2009}).

\section{Monte Carlo Study} \label{sec:mc}

We verify that we are able to recover a distribution described by a particular set of (hyper)parameters using a Monte Carlo injection study.
We create a simulated universe in which the black hole mass distribution follows our model with (hyper)parameters given in Tab.~\ref{table:prior}.
For simplicity, we draw all of the extrinsic parameters from the geometrically determined prior distribution used by \texttt{LALInference} (\cite{PhysRevD.91.042003}) with luminosity distance extending to $\unit{4}{Gpc}$.
We draw the masses according to Eq.~\ref{eq:model} and draw black hole spins uniformly in spin magnitude and isotropically in orientation.

Motivated by the prediction of pulsational pair-instability supernovae, our simulated universe includes a Gaussian component centred at the upper limit of the power law component, $\mx=\msol{35}$.
The Gaussian component has a width $\spp=\msol{1}$ and the mixing fraction $\lam=0.1$.
The inferred value of $\al$ is covariant with other parameters of the model, e.g., for the first four detections, decreasing the maximum black hole mass, decreases the inferred value of $\al$ (\cite{Fishbach17b}).
We set $\al=1.5$ for our injection study, which is consistent with that analysis.
We choose the spectral index of the secondary mass distribution to be $\bt=2$ to reflect the preference of population synthesis models to produce near equal mass binaries.
We impose a lower mass cut-off of $\msol{3}$ with a turn-on of $\dm = \msol{5}$.
These values are chosen to be consistent with current observational data.
The distribution of primary masses with this choice of (hyper)parameters is shown in Fig.~\ref{fig:model}.
We can see that our model gives us a bimodal distribution with peaks at $\approx\msol{7}$, due to the smooth turn-on, and $\msol{35}$, due to PPSN.

\begin{figure}
\includegraphics[width=\linewidth]{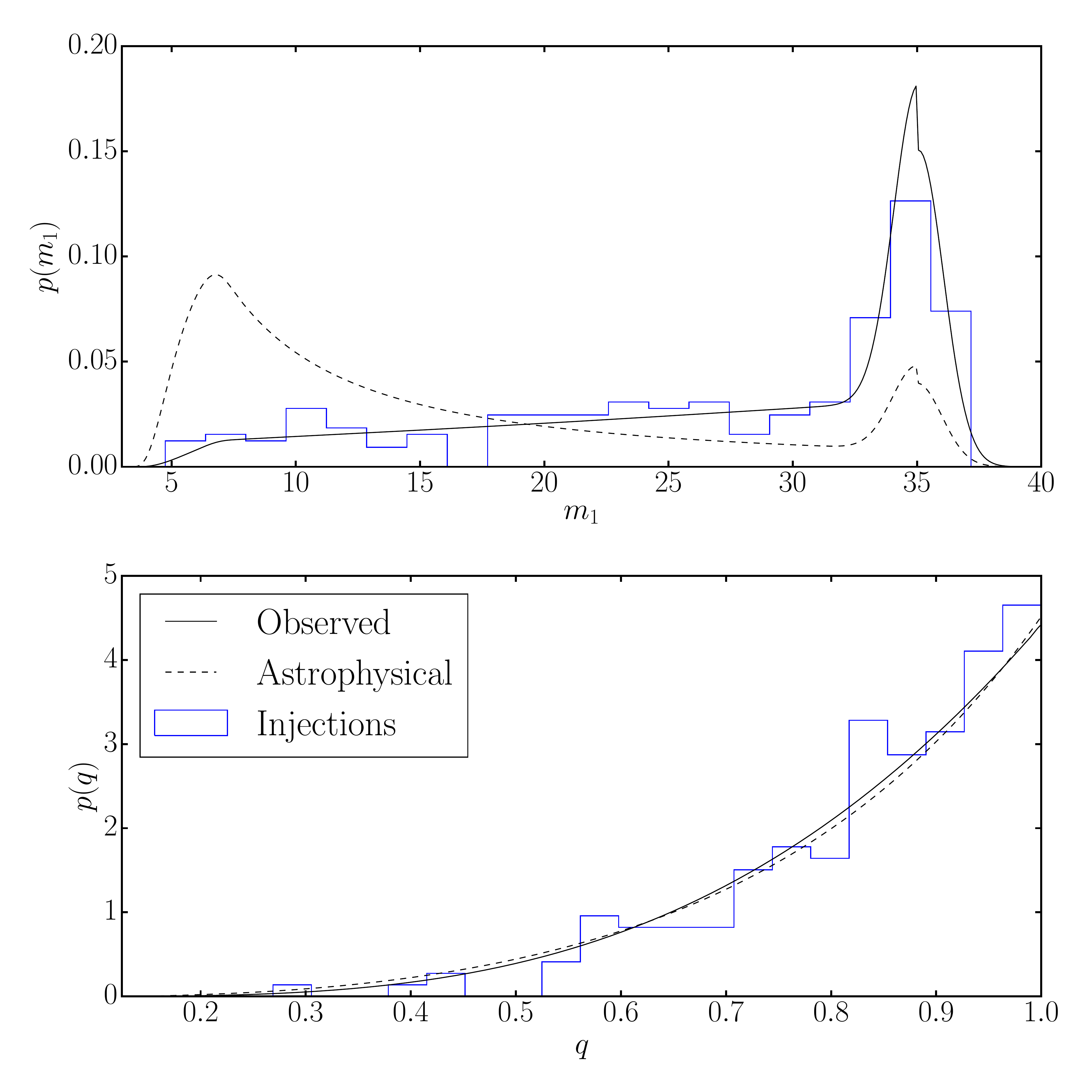}
\caption{
The distribution of source-frame primary mass and mass ratio ($q\equiv m_{2}/m_{1}$) for our simulated universe, see Tab.~\ref{table:prior}.
The dashed and solid lines show the distribution before and after accounting for selection biases respectively.
The blue histogram indicates the injected values.}
\label{fig:universes}
\end{figure}

To enforce selection effects, we keep only binaries with optimal matched filter signal to noise ratio, $\rho>8$, in a single Advanced LIGO detector operating at design sensitivity (\cite{Abbott2016}).
We generate a set of 200 events for our simulated universe.
Each signal is then injected into a three detector LIGO-Virgo network with all detectors operating at their design sensitivities.
Fig.~\ref{fig:universes} shows the distribution of primary masses and mass ratio in our simulated universe before (dashed) and after (solid) accounting for observation bias.
The blue histogram indicates the injected values.

Using the recovered posterior distributions for the injected events, we employ the statistical methods described above for each of our models.
The Bayes factors comparing $H_5$ to the others are enumerated in Table~\ref{table:sim-ev}.
In Table~\ref{table:sim-ev} we also give an approximate number of events needed to reach our threshold $\ln BF=8$, assuming linear growth of $\ln BF$ with number of detections.
We consider two cases.
``Cosmic'' assumes zero measurement error.
All uncertainty comes from cosmic variance.
``Design'' uses posterior samples obtained through running \texttt{LALInference} for a three detector network operating at design sensitivity.
Including measurement errors reduces our resolving power between any pair of models by a significant factor for all the models.
Unless otherwise specified we will refer to the Design Bayes factors.
Below, we consider the effect of each of the modifications on the mass distribution model.

\begin{table}[t]
\begin{ruledtabular}
\begin{tabular}{cccccc}
& $H_{0}$ & $H_{1}$ & $H_{2}$ & $H_{3}$ & $H_{4}$ \\ \hline
Cosmic $\ln BF^{5}_{i}$ & 253.0 & 55.0 & 18.0 & 31.0 & 1.0 \\
Design $\ln BF^{5}_{i}$ & 161.0 & 14.0 & 5.0 & 7.0 & -1.0 \\ \hline
$N_{\text{expected}}$ & 10 & 100 & 300 & 250 & $ \gg 200$ \\
\end{tabular}
\end{ruledtabular}
\caption{
The log Bayes factor comparing each of the hypotheses summarized in Tab.~\ref{table:prior} to the correct model, $H_{5}$, given the 200 samples shown in Fig.~\ref{fig:universes} with population (hyper)parameters as specified in Tab.~\ref{table:prior}.
``Cosmic'' indicates that the masses are used with no measurement error, this represents an upper limit on how well we can differentiate the two distributions.
``Design'' uses the output of LALInference for a three detector Advanced LIGO/Virgo network operating at design sensitivity.
The bottom row gives an approximate number of events to reach our threshold of Design $\ln BF=8$.
Measuring the shape of the low-mass cut-off will require many detections, due to the lower sensitivity at low masses.
}
\label{table:sim-ev}
\end{table}

\subsection{Upper-Mass Cut-Off}

After 200 events, our model without the variable upper-mass cut-off, $H_0$, is disfavored with a log Bayes factor of $\sim160$.
We determine how many events are necessary to surpass the threshold of $\ln BF^{5}_{0}=8$ by considering subsets of our injection set.
After 20 detections $\ln BF^{5}_{0} \sim N(\mu=13.3,\sigma=2.4$).
Here, $N(\mu, \sigma)$ denotes a normal distribution with mean $\mu$ and variance $\sigma^2$.

Given this, we expect to be able to identify an upper-mass cut-off in the mass distribution after $\lesssim20$ events.
We note that $\mu$ grows linearly with number of events, this scaling is used in Table~\ref{table:sim-ev} to approximate the number of events to reach $\ln BF=8$.
This is consistent with a similar study by~\cite{Fishbach17b}.

\subsection{PPSN Peak}

\begin{figure}
\includegraphics[width=\linewidth]{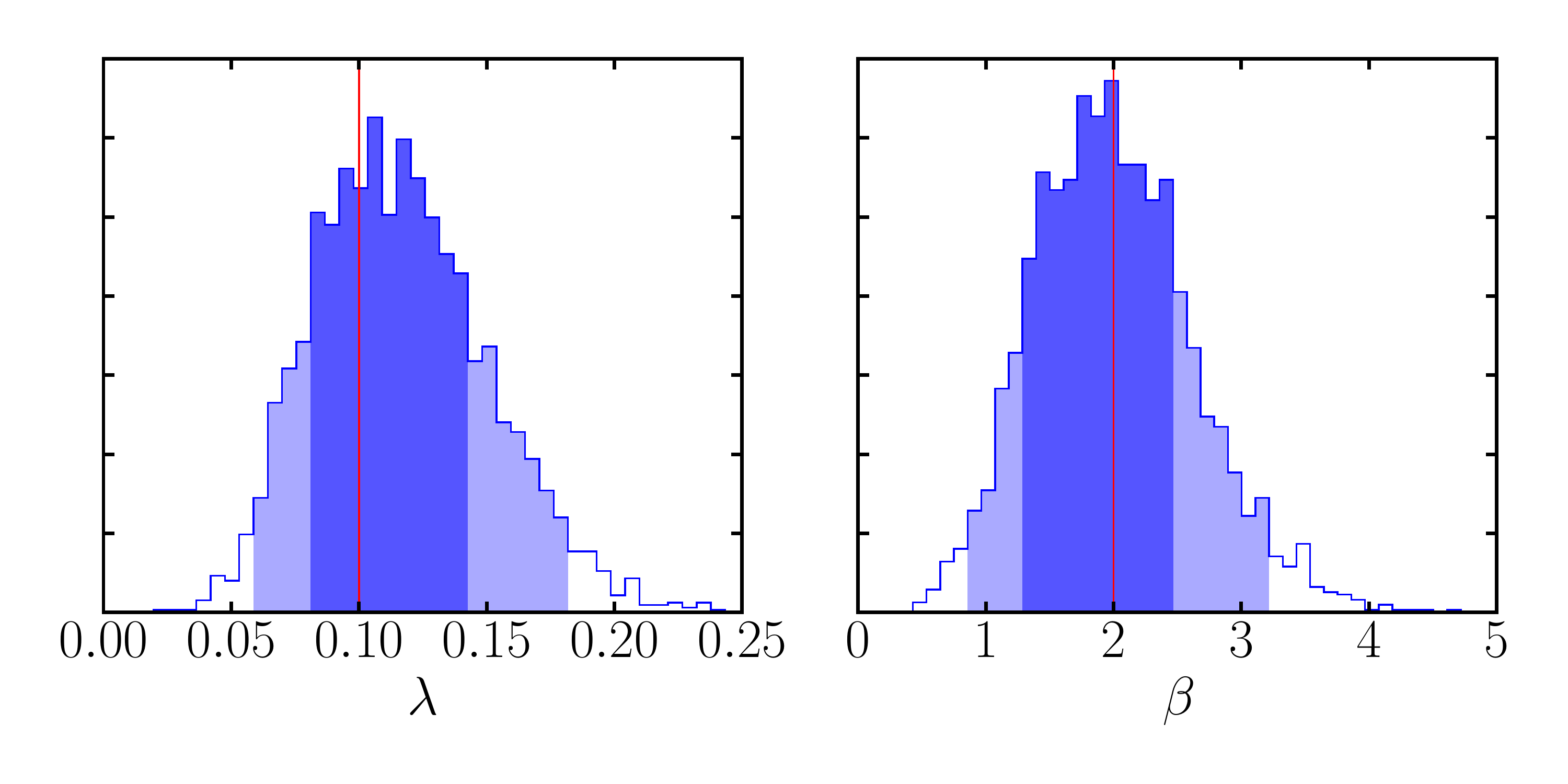}
\caption{
The posterior on the fraction of black holes formed through PPSN and the power-law index on the mass ratio using our 200 events injected into Gaussian noise.
We can measure the fraction of black holes formed through PPSN to be $\lam\sim0.11^{+0.07}_{-0.04}$ at 95\% confidence.
We can determine the spectral index of the mass ratio distribution to within $\pm 1$ at 95\% confidence with 200 detections.
}
\label{fig:lam-bt-post}
\end{figure}

After 200 events, the posterior distribution on $\lam$, the fraction of black holes formed through PPSN, is shown in Figure~\ref{fig:lam-bt-post}.
We measure the maximum posterior probability point and 95\% highest density confidence interval (HDI) to be $\lam\sim0.11^{+0.07}_{-0.04}$ (all future confidence regions will be 95\% HDI unless specified) and disfavor $\lam=0$ at $\gtrsim 3 \sigma$.
Correspondingly, $\ln BF^{5}_{2} = 4.9$ which is moderate evidence for the existence of the PPSN peak, but below our threshold for a confident detection.

\begin{figure}
\includegraphics[width=\linewidth]{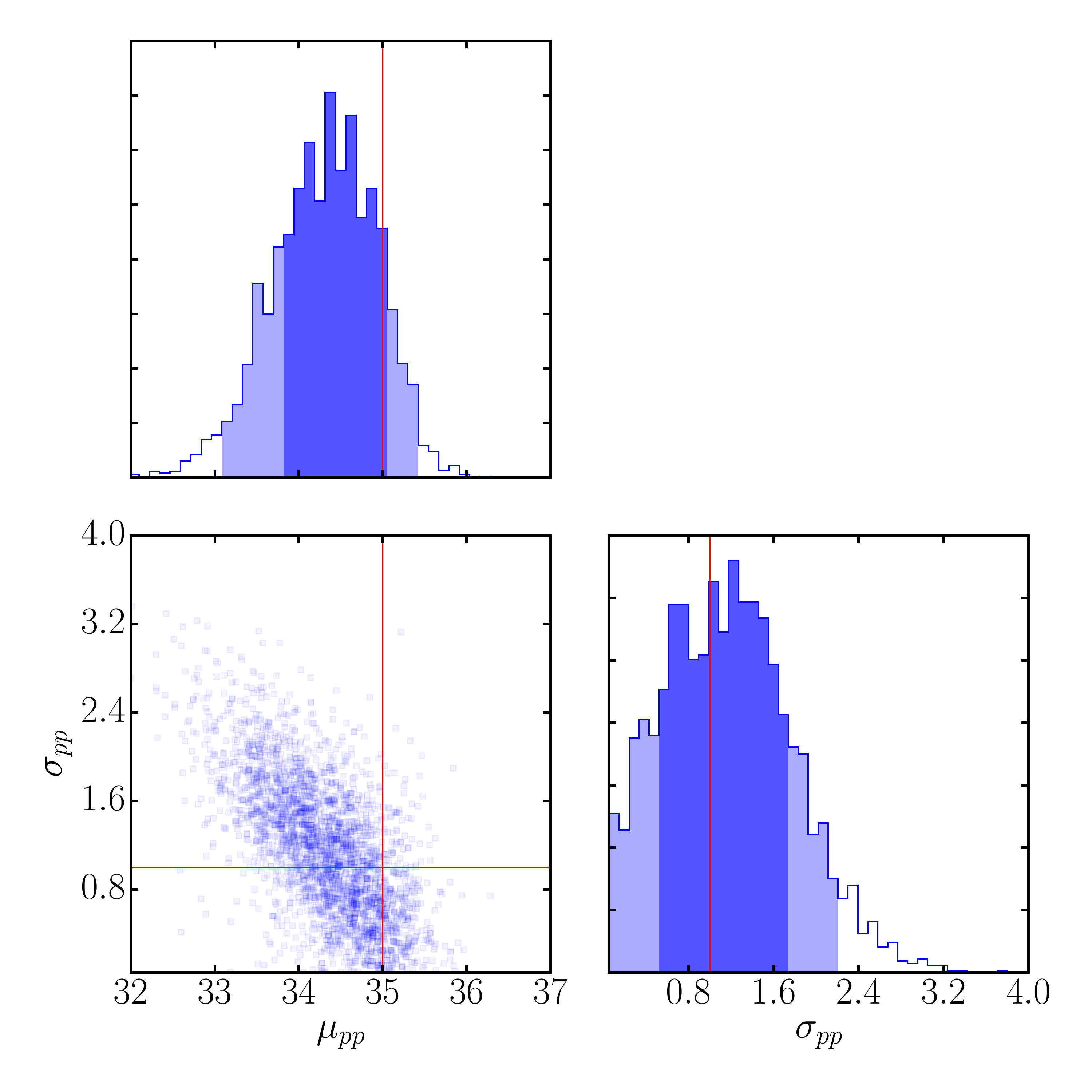}
\caption{
The posterior on the mean and width of the PPSN peak using our 200 events injected into Gaussian noise.
After 200 detections we can measure the position and width of the PPSN peak to within $\sim \msol{1}$ at 95\% confidence.
The dark and light shaded regions indicate the one-dimensional 68\% and 95\% confidence intervals.
}
\label{fig:pp-post}
\end{figure}

Figure~\ref{fig:pp-post} shows the one and two-dimensional posterior distribution for the position and width of the PPSN peak.
We meausre $\mpp = 34.4^{+1.0}_{-1.2} M_{\odot}$, $\spp = 1.2^{+0.9}_{-1.2} M_{\odot}$.
We note that the peak and width of the distribution are covariant, with smaller values of $\mpp$ requiring a larger $\spp$.
This is unsurprising as the highest mass black holes, $\sim \msol{40}$, must be accounted for.
The posterior distribution on $\lam$ shows no significant correlation with $\mpp$ or $\spp$.

\subsection{Mass Ratio}

The posterior distribution on $\bt$ is shown in Figure~\ref{fig:lam-bt-post}.
After 200 events, the $1\sigma$ and $2\sigma$ confidence intervals on $\bt$ span $1.1$ and $2.2$ respectively.
We disfavor $H_3$ with a log Bayes factor of $\ln BF^{5}_{3}=7.1$ after our 200 injections, just below our threshold of 8.

\subsection{Low Mass}

\begin{figure}
\includegraphics[width=\linewidth]{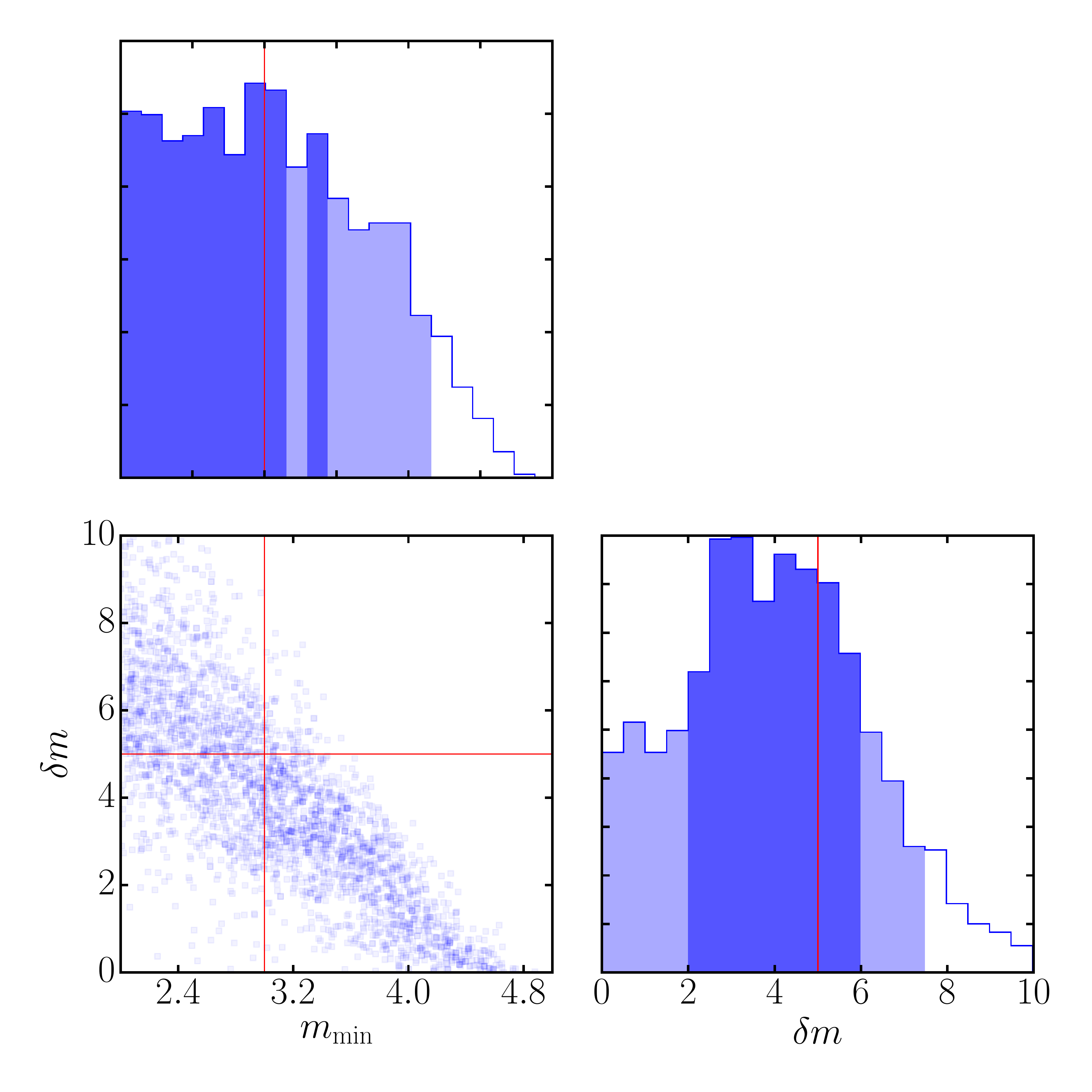}
\caption{
The posterior on the (hyper)parameters describing the low-mass end of the black holes mass spectrum using our 200 events injected into Gaussian noise.
These parameters are difficult to measure as only about 5\% of events have $m_1<\msol{8}$.
We can see the clear covariance between a low minimum mass and a long turn-on length.
}
\label{fig:mndm-post}
\end{figure}

The (hyper)parameters describing the low-mass end of the distribution are more difficult to measure than the high-mass (hyper)parameters due to the observation bias favoring high-mass systems.
After 200 events, there is no evidence for or against the low-mass smoothing described by $\dm$.
This is unsurprising since only 9 injected binaries have $m_1\lesssim\msol{8}$, above which $H_{4}$ and $H_{5}$ are identical.
Measuring the same events with improved strain sensitivity would not improve our sensitivity to $\dm$.
We are limited by cosmic variance: Cosmic $\ln BF^{5}_{4}=1$.
The two-dimensional posterior distribution on $\mn$ and $\dm$, Figure~\ref{fig:mndm-post}, shows the correlation between low minimum masses and long turn-on lengths.

\subsection{Mass Distribution Recovery}

\begin{figure}
    \includegraphics[width=\linewidth]{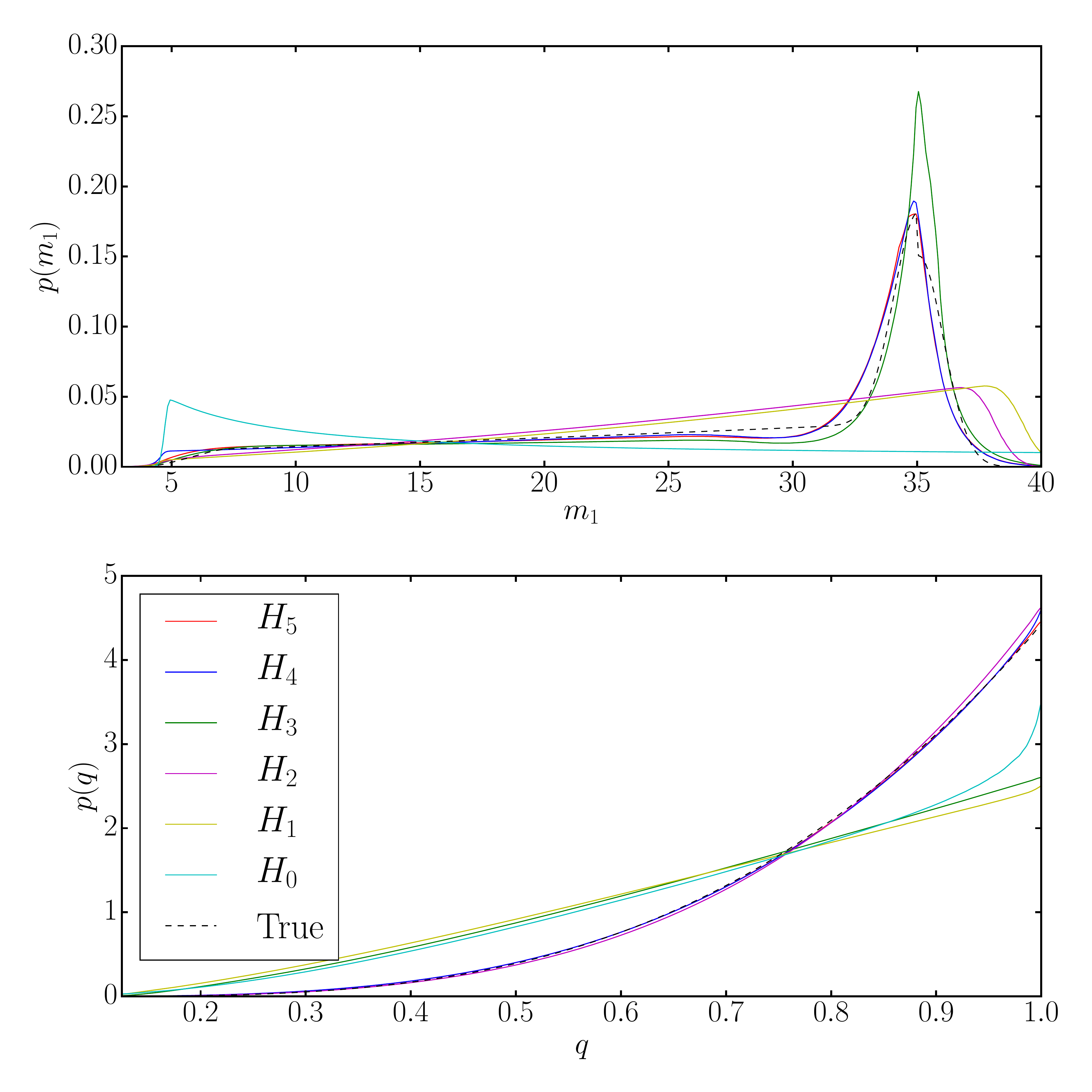}
\caption{
Posterior predicitive distribution (see Eq.~\ref{eq:ppd}) for $m_{1}$ and $q$ for our models from Tab.~\ref{table:prior} after 200 injected events.
The dashed black line indicates the true distribution and the notches indicate the injected values.
These distributions represent the observed distribution of source-frame masses.
}
\label{fig:ppd}
\end{figure}

As a qualitative measure of the difference between the inferred mass distributions, we plot the posterior predictive distribution for the primary mass given the binaries in our injection study for our models in figure~\ref{fig:ppd}.
The dashed black line indicates the injected distribution.
We can see the effect of the different (hyper)parameterizations.

In order to accommodate the lack of black holes with $m\gtrsim \msol{40}$, $\al$ is overestimated in model $H_{0}$. This leads to an overly steep inferred distibribution and an overestimate of the total merger rate (see Sec.~\ref{sec:rate}).
Models $H_{1}$ and $H_{2}$, which do not include the Gaussian component, favor a mass spectrum which is less steep than the injected distribution.
This manifests as a positive gradient after accounting for observation bias.
If we do not fit the mass ratio power-law index, we overestimate the number of heavy black holes.
This bias is seen as an overestimate of $\lam$ in $H_{3}$ and maximum mass in $H_{1}$.

\subsection{Impact on the Merger Rate} \label{sec:rate}

The majority of binary black hole mergers are not individually resolvable by Advanced LIGO/Virgo.
Using a (hyper)parameterization which does not accurately describe the true distribution leads to a biased estimate of the fraction of mergers which are individually resolvable and hence the merger rate (\cite{Abadie2010,GW170104}).
Compact binary coalescences are a Poisson process which can be described by a merger rate $R(\Lambda)$.
For a detector with time-independent sensitivity and a model of the distribution of binary black hole systems, the merger rate can be inferred from: the number of observed events $N$, the sensitive volume of our detectors $V(\Lambda)$, and the observation time $T$,
\begin{equation}
    R(\Lambda) = \frac{N}{V(\Lambda) T},
\end{equation}
where
\begin{equation}
    V(\Lambda) = \int d\Theta \, \pi\left(\Theta \middle|\Lambda\right) V_{obs}(\Theta),
\end{equation}
and $V_{obs}(\Theta)$ is the sensitive volume to a given binary introduced in section Sec.~\ref{sec:inference}.

\begin{figure}
\includegraphics[width=\linewidth]{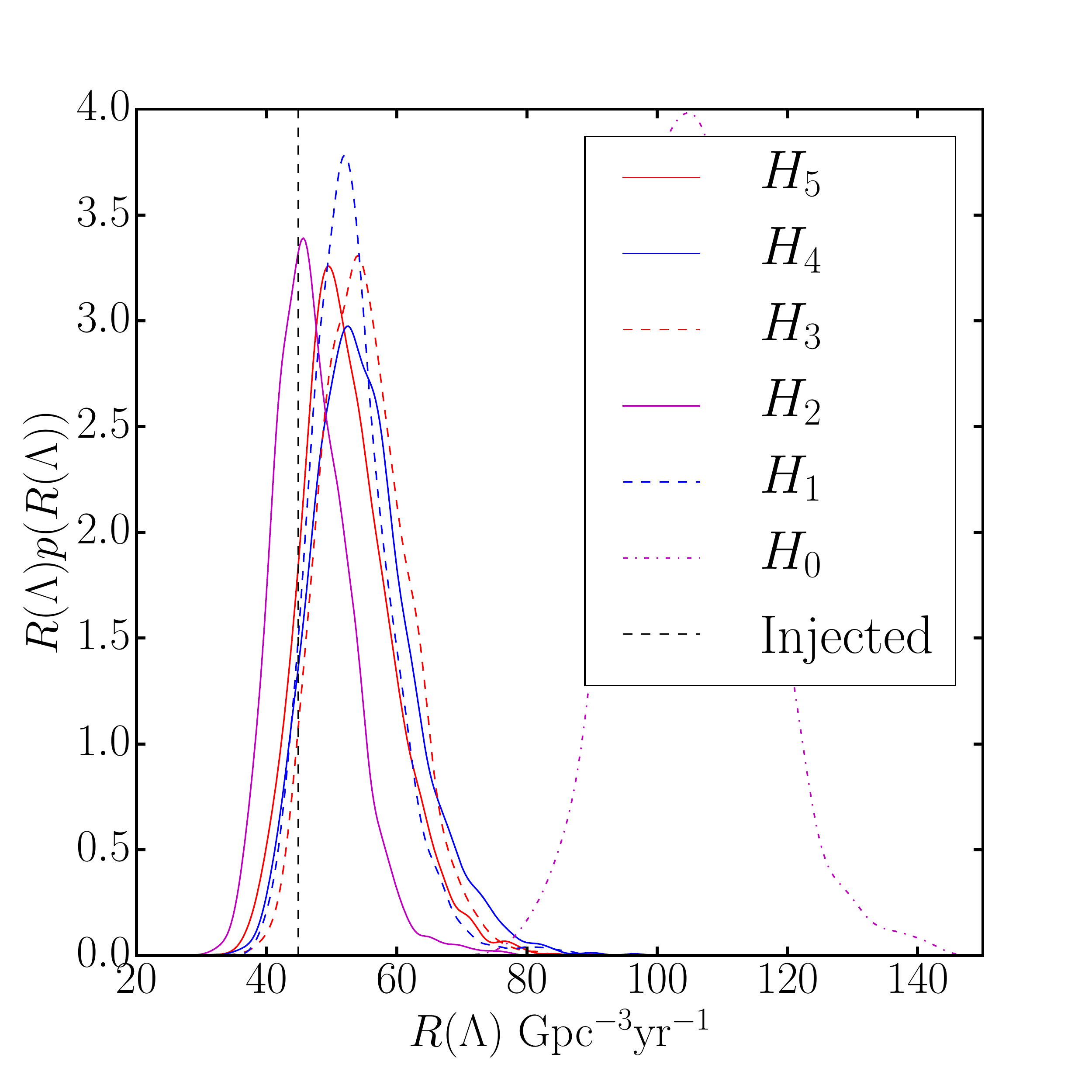}
\caption{
Posterior distribution for the binary black hole merger rate after 200 simulated events.
We assume a detection rate consistent with Advanced LIGO's first observing run, $N/T\approx\unit{23}{\text{yr}^{-1}}$, and ignore Poisson uncertainties.
The models are described in Tab.~\ref{table:prior}.
The dashed black line indicates the rate for the injected distribution.
If we do not fit the maximum mass (the dash-dotted line) the rate is overestimated by a factor of 2-3.
The inferred merger rate is not strongly sensitive to any of the other modifications to the mass function.
}
\label{fig:rates}
\end{figure}

To illustrate the dependence of $R$ on the mass distribution model, we calculate the posterior distribution for the inferred merger rate estimate for the models described in Tab.~\ref{table:prior}.
For each model, we compute the posterior distribution for $R$.
During the first observing run of Advanced LIGO, $\sim 3$ binary black hole mergers were identified in $\sim 48$ days joint observing time (\cite{PhysRevX.6.041015}).
We use these values, $N=3$, $T=\unit{(48/365)}{\text{ yr}}$ to normalize our rate estimates.
We neglect the (currently large) Poisson uncertainty in the arrival rate since this will be small once 200 detections have been made.
Fig.~\ref{fig:rates} shows the posterior distribution for the merger rate for each of our models.
We note that if we assume the power law mass distribution extends out to $\msol{100}$, the dash-dotted line, we overestimate the merger rate by a factor of 2-3.
This is due to the much larger $\al$ required to be consistent with the lack of detections at high masses.
A similar result is obtained in~\cite{WysockiThesis} by simultaneously fitting the merger rate and power-law spectral index.

\subsection{Impact on the Stochastic Background}

\begin{figure}
\includegraphics[width=\linewidth]{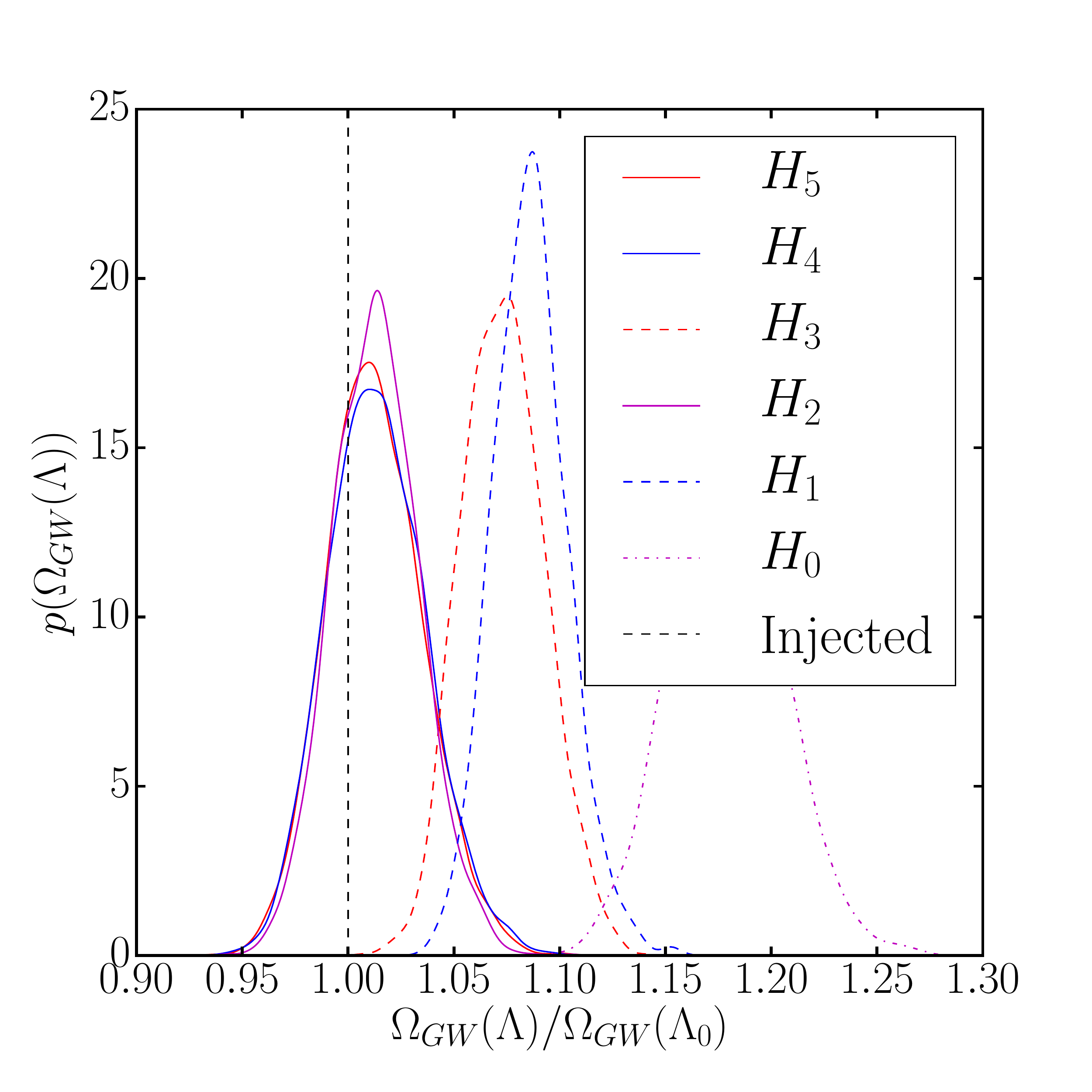}
\caption{
Posterior distribution for the ratio of the amplitude of the expected stochastic gravitational-wave background to the value using the injected distribution.
The models are described in Tab.~\ref{table:prior}.
The dashed lines indicate models in which the mass ratio distribution is assumed to be uniform.
The dash-dotted line indicates the model in which the maximum black hole mass is fixed to be $\msol{100}$.
Allowing the maximum mass of the power law component to vary decreases the predicted amplitude of the stochastic background by $\sim 10\%$.
Relaxing the assumption that the distribution of secondary masses in uniform between the minimum mass and the primary mass decreases the predicted amplitude by $\sim 10\%$.
}
\label{fig:sgwb}
\end{figure}

Unresolvable mergers are widely believed to make the dominant contribution to the SGWB (\cite{GW170817_stoch,gw150914_stoch,o1_iso}).
The SGWB is typically characterized by the ratio of the energy density of the universe in gravitational-waves to the energy required to close the universe, $\Omega_{GW}$.
The most sensitive frequency of current detectors to the SGWB is $\sim \unit{30}{\text{Hz}}$, this frequency corresponds to the inspiral phase of all binaries relevant to this work.
The energy density due to binary black hole mergers depends on the distribution of chirp masses and the merger rate (\cite{Zhu11}),
\begin{equation}
\Omega_{GW} \sim \langle \mathcal{M}^{5/3} \rangle R.
\end{equation}

As seen above, cutting off the mass distribution around $\msol{40}$ leads to a reduction in the merger rate, however, this is accompanied by an increase in $\mathcal{M}^{5/3}$.
Overall, this leads to an $\sim 10\%$ reduction in the expected SGWB as seen in Figure~\ref{fig:sgwb}.
We also observe that relaxing the assumption that the secondary mass is uniformly distributed leads to a further $\sim 10\%$ reduction in $\Omega_{\text{GW}}$.
This is because the chirp mass is maximized for equal mass binaries for a given primary mass.
These reductions are smaller than the current uncertainty on the amplitude of the background due to Poisson uncertainty in the observed merger rate.
The current method of searching for this background is by cross-correlating the strain data from the two LIGO detectors, this method will take more time to resolve a weaker background.

The cross-correlation method is expected to require years of observation before the background can be resolved.
Recently, a method involving searching directly for the stochastic background due to binary black hole mergers has been introduced in \cite{Smith2017}.
This method is expected to be able to detect this component of the background using days of data.
Since this method relies on the rate of binary black hole mergers rather than $\Omega_{\text{GW}}$ it will be more sensitive to the black hole mass function than cross-correlation searches.

\section{Discussion} \label{sec:discussion}

The first gravitational-wave detections are revealing a previously unexplored population of black holes.
While we are still in the regime of small-number statistics, the systems observed to date may be suggestive of a cut-off in the black hole mass spectrum at $\sim\msol{40}$.
This is consistent with the predicted black hole mass distribution if stars with initial masses $\mzams\gtrsim\msol{100}$ undergo pulsational pair-instability supernovae.
We hypothesize that, if this is the cause of the cut-off, then there should be a corresponding excess of black holes at around the same mass.
We construct a phenomenological model, which captures this behavior.
In agreement with~\cite{Fishbach17b}, we find that the presence of an upper mass cut-off can be identified at high significance with $O(10)$ events.

We highlight several other interesting results that can be obtained using 200 detections at design sensitivity:
\begin{enumerate}
\item We will be able to identify the presence of an excess due to PPSN at $\sim3\sigma$ and constrain the fraction of black holes forming through PPSN to within $\sim 0.05$ at 95\% confidence.
\item We can measure the position and width of the PPSN graveyard to within $\sim \msol{1}$.
\item We will be able to measure the power-law index on the mass ratio to within $\sim \pm 1$.
\end{enumerate}
Detailed measurement of the low-mass end of the mass distribution will most likely require 1000s of detections and may have to wait for future detectors, e.g., the proposed Einstein Telescope (\cite{ET}) or Cosmic Explorer (\cite{CosmicExplorer}).

We demonstrate that neglecting the presence of either a cut-off or a mass peak can lead to a mis-recovery of the astrophysical distribution of black holes in merging binaries.
For example, the higher sensitivity of current detectors to high-mass binaries means that in order to fit the upper mass range well, the low-mass distribution is biased.
This leads to incorrect estimates of the total binary black hole merger rate and the predicted amplitude of the SGWB.
The amplitude of the SGWB is also sensitive to the distribution of mass ratios.

Our analysis assumes that a clear distinction can be made between binary black hole systems and other compact binaries.
In reality, if there is not a well-defined mass gap between neutron stars and black holes, it will be non-trivial to distinguish between binary black hole, neutron star-black hole, and binary neutron star systems (\cite{Yang2017}).
Although, differences in, e.g., the spins of the component objects may enable this distinction (\cite{Littenberg2015}).
Our framework can be naturally expanded to include these other classes of compact binaries.

\acknowledgements
This is LIGO document P1700449-v5.
We would like to thank Richard O'Shaughnessy for helpful comments.
ET is supported by ARC CE170100004 and FT150100281.

\bibliography{mass.bib}

\end{document}